\documentclass[12pt,oneside,english]{amsart}
\usepackage{mathptmx}

\usepackage[T1]{fontenc}
\usepackage[latin9]{inputenc}
\usepackage{geometry}
\usepackage{amsthm}
\usepackage{graphicx}

\makeatletter
\numberwithin{equation}{section}
\numberwithin{figure}{section}
\theoremstyle{remark}
\newtheorem*{rem*}{\protect\remarkname}

\makeatother

\usepackage{babel}
\providecommand{\remarkname}{Remark}

\begin{document}
\title{the periodic table of the elements with $4n^{2}$ $n=2,3,\dots$ periods}
\author{G. Beylkin }
\address{Department of Applied Mathematics \\
 University of Colorado at Boulder \\
 526 UCB \\
 Boulder, CO 80309-0526 }

\begin{abstract}
A modification of the standard periodic table of the elements reveals
$4n^{2}$ periods, where $n=2,3,\dots$. The new arrangement places
hydrogen with halogens and keeps the rare-earth elements in the table
proper (without separating them as they are in the standard table).
Effectively, periods in the modified table are defined by the halogens
rather than by the noble gases. The graph of ionization energy of the
elements is presented for comparison of periods in the standard and
the modified tables.
\end{abstract}

\maketitle

\section{Introduction}

Since the periodic table of the elements was first published by Mendeleev
in 1869, there were many attempts to find the ``most natural'' representation
of the periodic law. These attempts to structure the periodic table
were numerous in the beginning of 20th century and are well documented
in \cite{SPRONS:1969}. The periodic law has been qualitatively explained
with the advent of Quantum Mechanics (QM) and, for many years now,
the periodic table remains virtually unchanged. The standard table
may be found in any Chemistry textbook (for many variants of the standard
layout see e.g. \cite{PT}). The qualitative QM explanation of the
periodic table uses the Schr\"odinger's equation with the central
potential and the Pauli's exclusion principle stating that two or
more identical fermions cannot occupy the same quantum state. Since
the eigenvalues of the Schr\"odinger's equation have multiplicity
$n^{2}$, $n=1,2\dots$, combined with the Pauli's exclusion principle,
it leads to periods of length $2n^{2}$ in the periodic table (see
a more detailed discussion later). In other words, in the standard
periodic table a period finishes with the filling of an electron shell
and, therefore, the standard table of the elements consists of periods
of increasing lengths between noble gases. Indeed, the first period
of the periodic table has two elements, followed by two periods of
$8$ elements, two periods of $18$ elements, a period of $32$ elements
and an incomplete period.

On the other hand, the eigenvalues of the relativistic Dirac's equation
with the central potential have multiplicity $2n^{2}$, $n=1,2\dots$,.
If we were to add two spin states of the nuclei, then the multiplicity
becomes $4n^{2}$. We note that the $4n^{2}$ law for the length of
periods was suggested early on by Rydberg (see comments in \cite{SPRONS:1969}).
However, if we attempt to build the periodic system starting with
$n=1$ (so that the first period has length $4$), then the resulting
table is inconsistent. 

In this note we show that by starting with $n=2$, the periods of
length $4n^{2}$ yield a simpler and a more symmetric form of the
periodic table. Consequently, we introduce a new form of the periodic
table of the elements which, due to the new definition of the period,
has symmetry and consistency superior to that of the standard table.
In the new form of the table a period \textit{finishes} one element
before a halogen and thus has a different meaning than in the standard
table. The periods in the new table are of length $4n^{2}$ (or twice
$2n^{2}$), where $n=2,3,\dots$. 

In the new table there is no need to separate the lanthanide and actinide
elements from the main table and hydrogen has its proper place among
halogens. The new table consists of six periods. It has two periods
of $8$ elements ($2\cdot2n^{2}$, $n=2$), two periods of $18$ elements,
($2\cdot2n^{2}$, $n=3$), and two periods of $32$ elements ($2\cdot2n^{2}$,
$n=4$). Remarkably, the new periodic table does not significantly
change the composition of the groups of elements. Only four groups
of the new periodic table partially differ in their arrangement from
that of the standard table. We discuss this difference below.

As additional evidence that the new arrangement of the elements adequately
incorporates properties of the elements, we present the ionization
curve which we split into segments according to the periods of the
new table and compare it with the corresponding segmentation of the
ionization curve according to the standard table.

\begin{figure}[h]
\begin{centering}
\includegraphics[bb=500bp 10bp 792bp 612bp,scale=0.5]{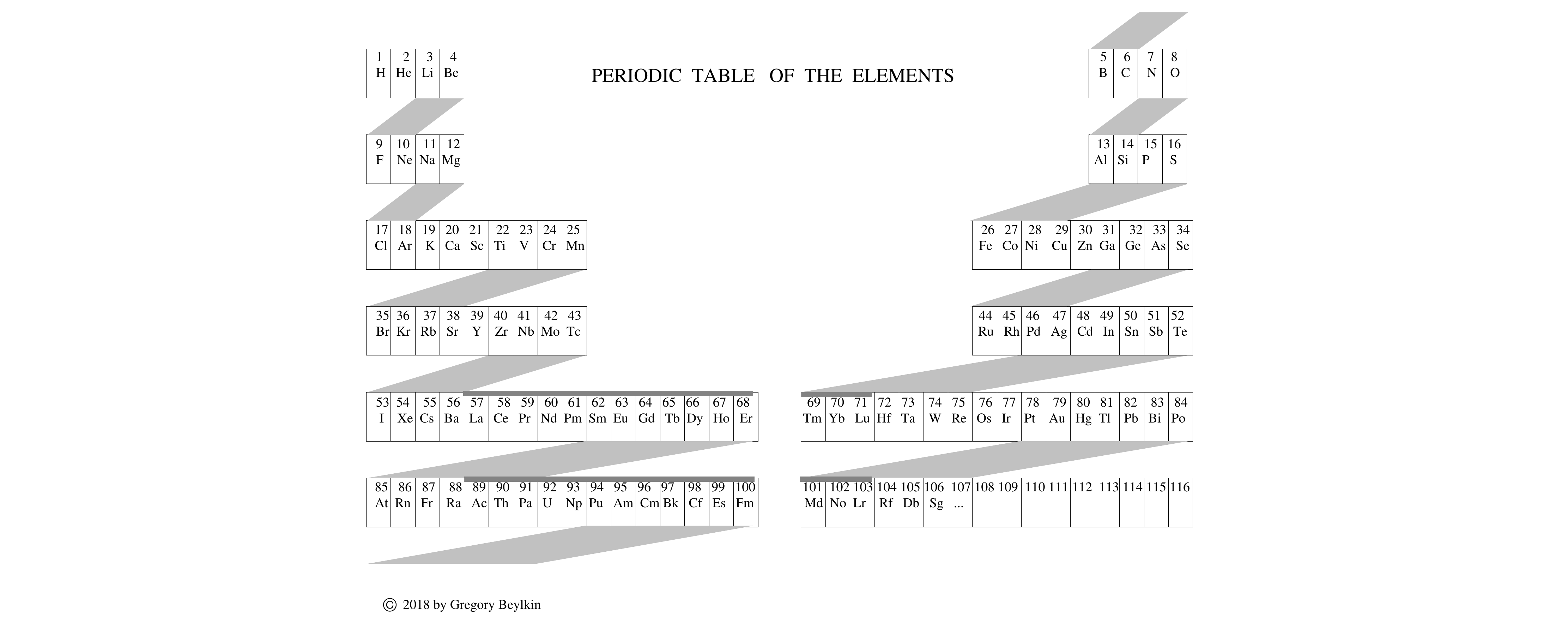}\caption{A new form of the periodic table of the elements obtained by folding
a strip with the elements arranged by their atomic numbers. The lanthanide
and the actinide elements of the standard table are highlighted. \label{fig:PT}}
\par\end{centering}
\end{figure}
\begin{figure}[h]
\centering{}\includegraphics[bb=500bp 50bp 792bp 400bp,scale=0.47]{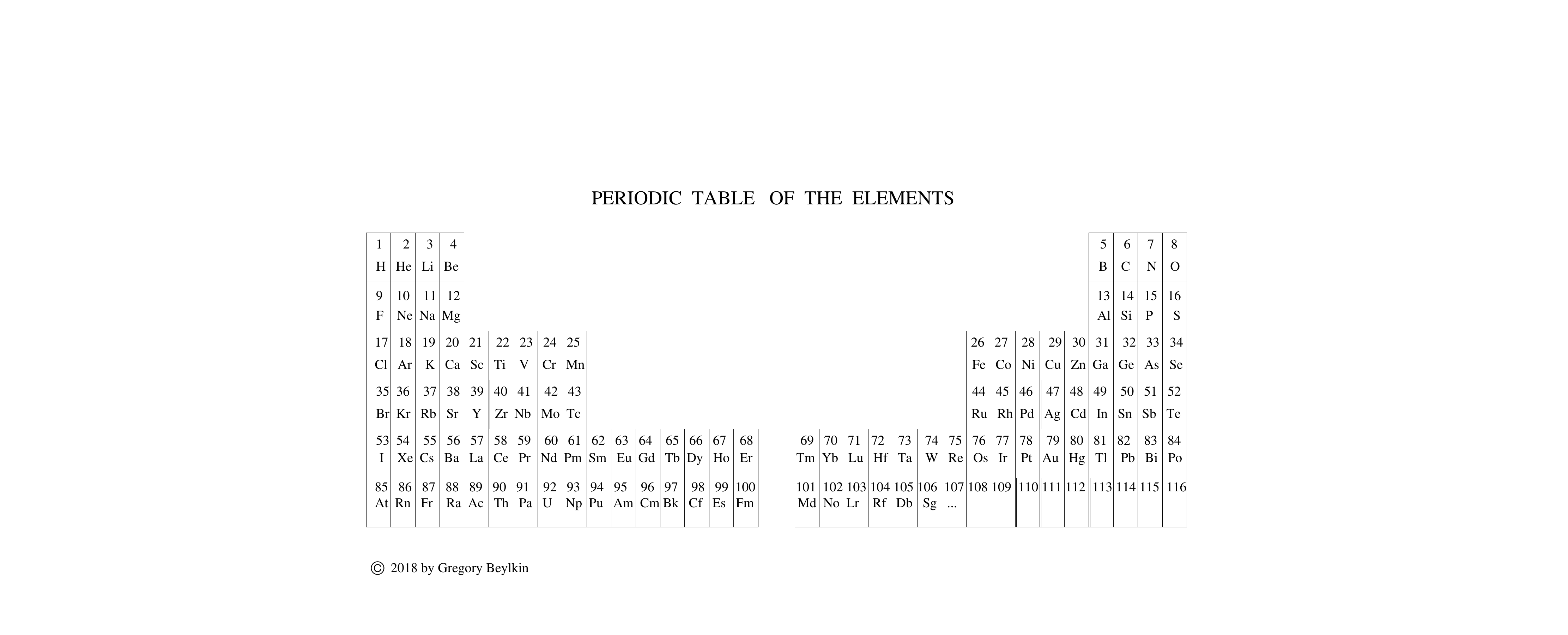}\caption{A new form of the periodic table of the elements.\label{fig:PT-1}}
\end{figure}

\section{Construction of the new periodic table}

Let us take a continuous strip of paper and, on one side of the strip,
write all the elements in the order of their atomic numbers. We then
form a spiral with the strip such that the two most chemically distinct
groups, the group of halogens (in which we include hydrogen) and the
group of noble gases, are properly aligned. By flattening the strip
on a plane and folding it in the middle, we obtain the new periodic
table depicted in Figure~\ref{fig:PT}. The gray sloping strips indicate
the back side of the strip when folded into a spiral. These gray strips
are present in Figure~\ref{fig:PT} to illustrate the way this table
was initially obtained and are removed in Figure~\ref{fig:PT-1}.
A quick examination confirms a highly symmetric form of the new table.
\begin{rem*}
\noindent The first element of the periodic table, hydrogen, can be
considered to be either an alkali metal or a halogen. In the standard
table hydrogen is an alkali metal and the periods finish one element
before the next alkali metal (i.e., noble gases are the last elements
of a period). Our construction treats hydrogen as a halogen and, therefore,
periods finish one element \textit{before} the next halogen. This
is the new definition of the period which allows us to arrive at the
construction in Figure~\ref{fig:PT}.
\end{rem*}
\noindent \textbf{Comparison of groups in the new and the standard
periodic tables}

An examination of groups of the elements in the new periodic table
in Figure~\ref{fig:PT-1} shows that all the groups are exactly the
same as in the standard table with the exception of a partial rearrangement
of four groups. These groups in the new table are: ${\bf (Ti,Zr,Ce,Th)}$,
${\bf (V,Nb,Pr,Pa)}$, ${\bf (Cr,Mo,Nd,U)}$ and ${\bf (Mn,Tc,Pm,Np)}$.
The corresponding groups in the standard table are: ${\bf (Ti,Zr,Hf,Rf)}$,
${\bf (V,Nb,Ta,Ha)}$, ${\bf (Cr,Mo,W)}$ and ${\bf (Mn,Tc,Re)}$.

We note, however, that these rearrangements are among the elements
which have many similar properties and, therefore, their grouping
is acceptable in more than one way. Moreover, in some older tables
these four groups were organized as in the new table in Figure~\ref{fig:PT-1}.
The reason for later regrouping of the rare-earth elements was the
difficulty of accommodating all of them in the main table. The solution
offered to that problem was to remove the lanthanide and the actinide
elements from the main table. This caused a rearrangement of the groups
in question leading to the current form of the standard table. 

In the new table in Figure~\ref{fig:PT-1} the lanthanide and the
actinide elements are a part of the main table and there is nothing
special in their classification. Besides the rare-earth element problem
resolved in this simple way, hydrogen (which is at the other end of
the periodic table) has its proper place among halogens.

To conclude, the new periodic system does not cause any problems with
classification of the elements into groups. Instead, it adds a great
amount of simplification since all the elements are now in the table
proper.

\section{\textbf{A pseudo-periodicity of the ionization curve and its relation
to the new periodic system}}

\begin{figure}
\begin{centering}
\includegraphics[scale=0.6]{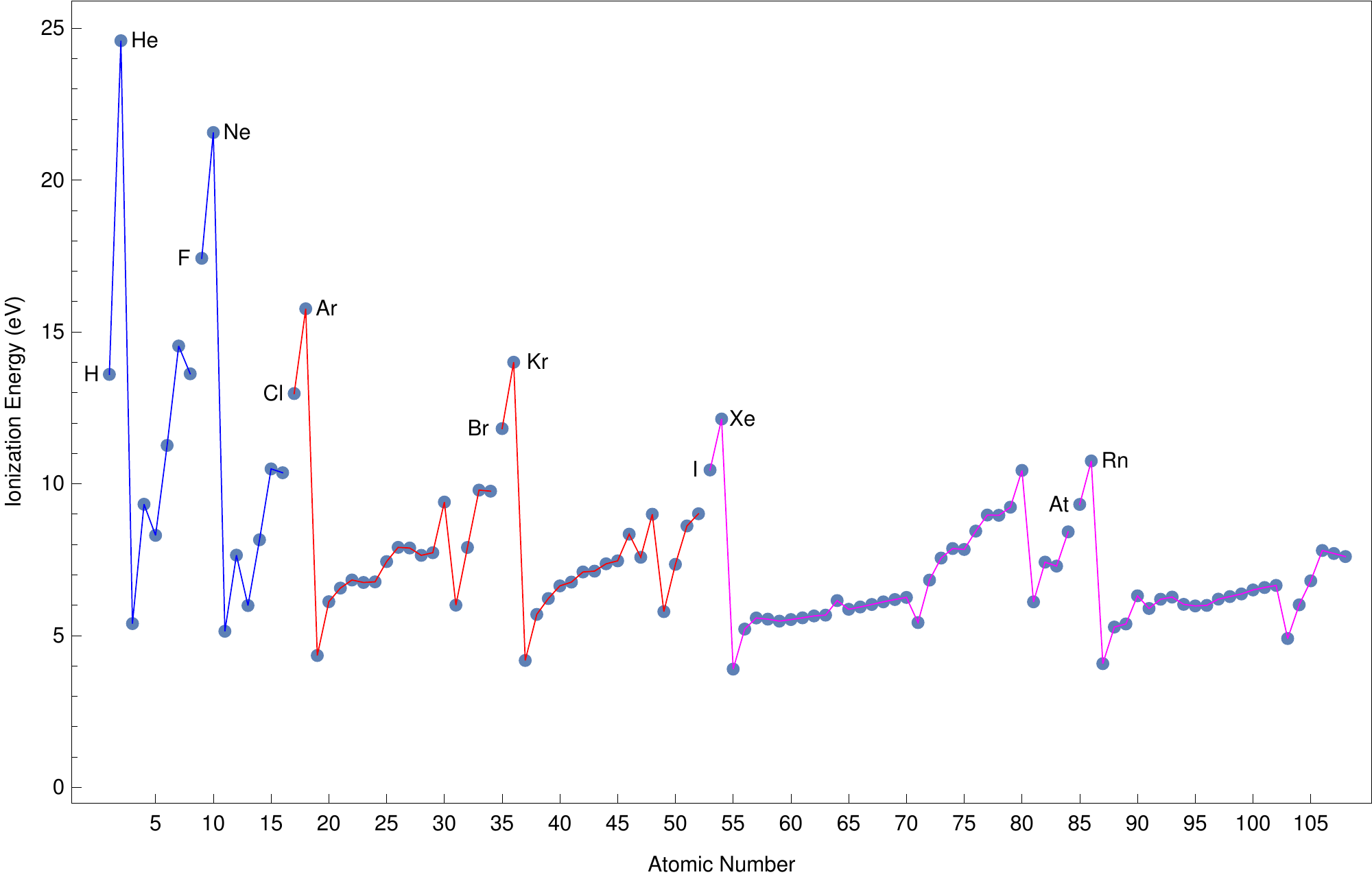}
\par\end{centering}
\begin{centering}
\includegraphics[scale=0.6]{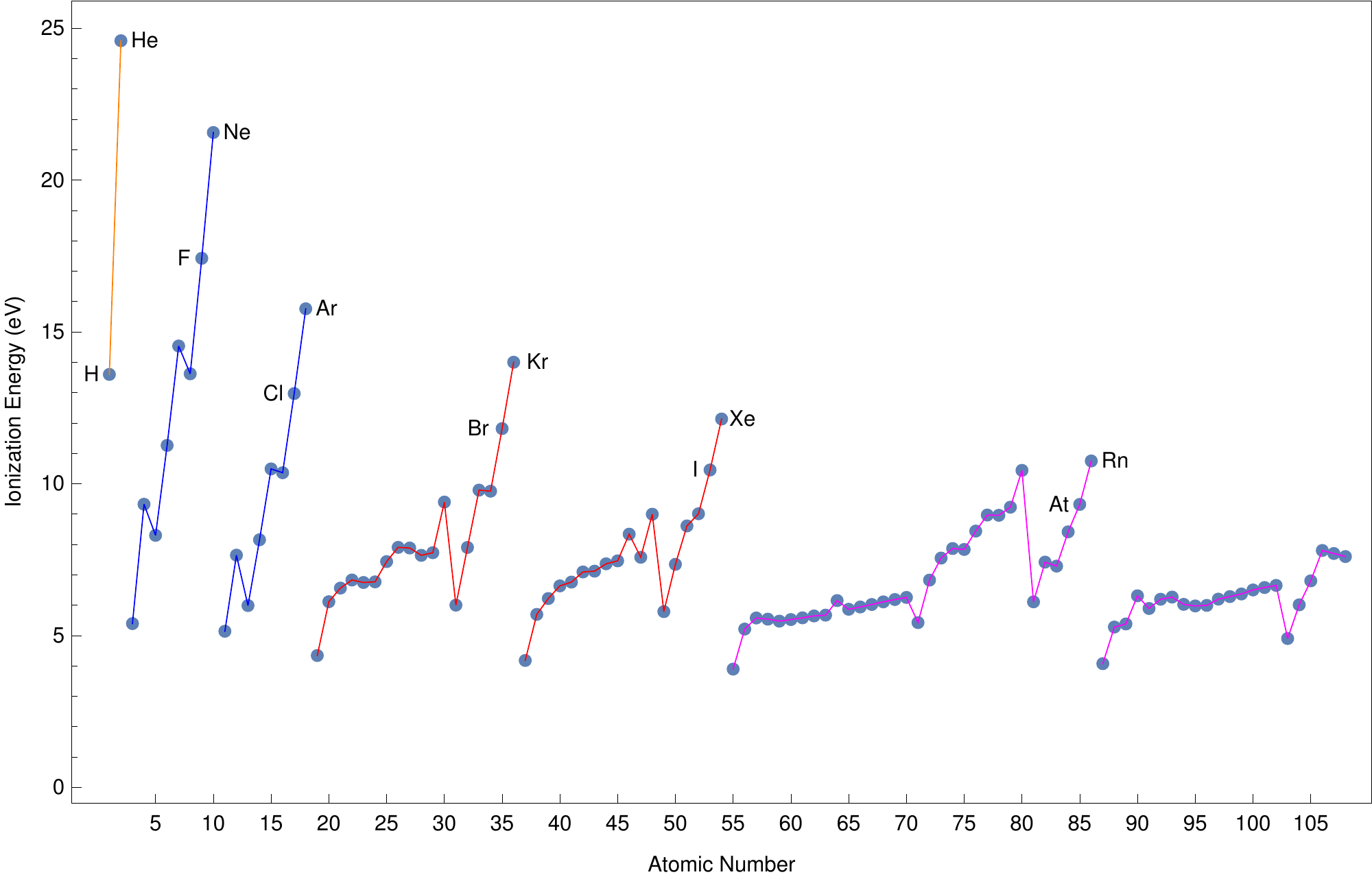}
\par\end{centering}
\begin{centering}
\caption{The pseudo-periodicity of the ionization energy of the elements is
highlighted using periods of the new table in Figure~\ref{fig:PT-1}
(top) and those of the standard table (bottom).\label{fig:IOC}}
\par\end{centering}
\end{figure}

The fact that the ionization curve has pseudo-periodic properties
was always given as a supporting fact for the periodic recurrence
of the properties of the elements. Subdivision of the ionization curve
into the segments according with the new definition of a period reveals
similarity between all of the segments. This is illustrated in Figure~\ref{fig:IOC}
where we have highlighted the subdivision of the curve according to
the periods in the new table and, for comparison, show the subdivision
according to the periods in the standard table.

\section{\textbf{On mathematical models for hydrogen and atomic structure.}}

The models for hydrogen and the atomic structure of the elements are
both based on the mathematical model of a charged particle in the
central field. Specifically, the eigenvalues of the Schr\"odinger's
equation with the central potential together with the added requirement
for the wave function to be antisymmetric, produce eigenvalues with
multiplicity $2n^{2}$, where $n=1,2,\dots$ is the principal quantum
number. Since the elements are multi-electron systems for which we
currently cannot solve the multiparticle Schr\"odinger's equation
directly, simpler models have been devised to evaluate electron configuration
of atoms. For example, the Hartree-Fock (H-F) equations are arrived
at starting from the multiparticle Schr\"odinger's equation by assuming
a separable approximation and enforcing antisymmetry by using the
Slater determinant as the wave function. The resulting effective potential
(which is now evaluated in the process of solving H-F equations) is
no longer spherically symmetric and eigenvalues split if compared
with the corresponding model of a charged particle in the central
field. Consequently, the electron structure can then be described
by introducing a shell model which qualitatively explains the standard
periodic system of the elements leading to periods of length $2n^{2}$. 

The experimental evidence shows that hydrogen spectrum can be split
(the so-called fine and superfine structures), so that the actual
multiplicity of eigenvalues in a model of two charged particles should
be $4n^{2}$. Such multiplicity of the spectrum can be arrived at
by considering the spin of proton and using the exclusion principle
for the wave function to account for the spins of both, electron and
proton. This bring us to the relativistic Dirac's equation for a charged
particle with spin which can be solved exactly yielding $2n^{2}$
energy levels for a fixed principal quantum number $n$, where the
number of levels does not depend on the charge of the particle \cite{DIRAC:1981}.
However, there is no equation describing the complete interaction
between electron and proton. The superfine structure is obtained as
a perturbation to the energy levels that follow from the Dirac's equation.
Furthermore, turning to multi-electron atoms, there is no accepted
``multiparticle Dirac equation'' at this time and relativistic effects
are estimated as corrections to the solutions of non-relativistic
equations. 

The new form of the periodic system in Figure~\ref{fig:PT-1} is
interesting in that it explicitly reveals periods of length $4n^{2}$
which consist of two sub-periods of the length $2n^{2}$. If we apply
the Pauli's exclusion principle to both electrons and nucleons of
the elements and, therefore, require the full wave function to be
antisymmetric with respect to exchanges of (separately) electrons
and nucleons, then it is reasonable to expect the $4n^{2}$ periodicity.
While the standard table also has these periods, they are ``hidden''
by having the first period with just two elements, hydrogen and helium.
Note that hydrogen does not have a definite place in the standard
table since one can think of it having only one electron or, alternatively,
missing one electron to form a closed shell. In the new table it found
its place with halogens. At the same time, the rare earth elements
are not separated from the main table. 

Currently there is no firm quantitative basis for selecting the new
periodic table over the standard one, although it is easy to argue
that the new form of the periodic table has a simpler structure than
the standard table. Hopefully, this well structured arrangement will
challenge and stimulate the development of methods for quantitative
description of the elements and, in particular, further development
of relativistic multiparticle models of atomic structure.

\bibliographystyle{siam}

\begin{thebibliography}{1}

\bibitem{PT}
https://www.meta-synthesis.com/index.html.

\bibitem{DIRAC:1981}
{\sc P.~A.~M. Dirac}, {\em The principles of quantum mechanics}, no.~27, Oxford
  University Press, 1981.

\bibitem{SPRONS:1969}
{\sc J.~W. Spronsen}, {\em The Periodic System of Chemical Elements: A History
  of the First Hundred Years}, Elsevier, 1969.

\end{thebibliography}

\end{document}